\documentstyle[12pt]{article}
\begin{document}
\begin{center}{\large{\bf A Possible Origin of Dark Matter, Dark Energy, and Particle-Antiparticle Asymmetry}}
\end{center}
\vspace*{1.5cm}
\begin{center}
A. C. V. V. de Siqueira
$^{*}$ \\
Departamento de Educa\c{c}\~ao\\
Universidade Federal Rural de Pernambuco \\
52.171-900, Recife, PE, Brazil.\\
\end{center}
\vspace*{1.5cm}
\begin{center}{\bf Abstract}

In this paper we present a possible origin of dark matter and dark
energy from a solution of the Einstein's equation to a primordial
universe, which was presented in a previous paper. We also analyze
the Dirac's equation in this primordial universe and present the
possible origin of the particle-antiparticle asymmetry. We also
present ghost primordial particles as candidates to some quantum
vacuum constituents.
\end{center}

 \vspace{3cm}

${}^*$ E-mail: acvvs@ded.ufrpe.br
\newline

\newpage

\section{Introduction}
$         $

The Einstein's theory of  general relativity is still the best
theory to describe problems in astrophysics and cosmology. However,
more recent observations in these two areas are apparently difficult
to be explained by general relativity. It raises the possibility to
consider models involving membranes and parallel universes, dark
matter, dark energy, and cosmological constant to explain the
behavior of large scale structures like galaxies, clusters of
galaxies, and the universe. Some physicists and astronomers believe
that the Einstein's theory needs to be modified while a quantum
gravity theory is not developed \cite{1}, \cite{2}, \cite{3}.
Actually, the Einstein's theory of general relativity offers
enormous analytical difficulties for problems involving
non-gravitational fields. However, for the primordial universe,
nature presents symmetries that make possible a better knowledge .
In this paper we present a possible origin of dark matter and dark
energy from a solution of the Einstein's equation, which was
presented in a previous paper, \cite{4}. We also analyze the Dirac's
equation for ghost fermions in the primordial universe and conclude
that they can be some of the constituents of the quantum vacuum. We
also present a justification to particle-antiparticle asymmetry.
\newline
This paper is organized as follows. In Sec. $2$, we present a
primordial and spatially flat solution of the Einstein's equation
with a massive scalar Klein-Gordon field and a cosmological
constant. The Jacobi equation is presented for this solution and two
primordial forces are identified as dark matter and dark energy,
respectively. In Sec. $3$,  we solve the Dirac's equation in a
vielbein basis for ghost spinors. The PCT theorem is applied and the
particle-antiparticle asymmetry is analyzed.  We also consider the
possibility that the ghost fermions are some constituents of the
quantum vacuum. In Sec. $4$, we summarize and conclude the results
of this paper.
\newpage
\renewcommand{\theequation}{\thesection.\arabic{equation}}
\section{\bf Exact Solutions of the Einstein's Equation}
\setcounter{equation}{0}
 $         $
In a previous paper we obtained three solutions of the Einstein's
equation with a  Klein-Gordon field and a cosmological constant. In
this paper we are interested only in the spatially flat solution. In
the following we review part of the referred  paper and present new
results. The convention used in a local basis was \cite{4}
\begin{equation}
R^{\alpha}{}_{\mu \sigma \nu }=\partial_{\nu} \Gamma_{\mu \sigma
}^{\alpha}-\partial_{\sigma}\Gamma_{\mu \nu}^{\alpha}+\Gamma_{\mu
\sigma}^{\eta}\Gamma_{n \nu}^{\alpha}-\Gamma_{\mu \nu}^{\eta}\Gamma
_{\sigma \eta}^{\alpha}
\end{equation}
with  Ricci tensor
\begin{equation}
R_{\mu \nu}=R^{\alpha}{_{\mu \alpha \nu }}.
\end{equation}
For this convention we have the following Einstein`s equation, with
a cosmological constant $\Lambda $,
\begin{equation}
R_{\mu \nu }-\frac{1}{2}g_{\mu \nu }R+\Lambda g_{\mu
\nu}=-\frac{8\pi G}{ c^{2}}T_{\mu \nu}
\end{equation}
where $T_{\mu \nu}$ is the momentum-energy tensor of a massive
scalar field,
\begin{equation}
T_{\mu \nu }=2 \nabla_{\mu} \phi \nabla_{\nu} \phi -g_{\mu
\nu}\nabla^{\alpha } \phi \nabla_{\alpha}\phi +m^{2} g_{\mu \nu
}\phi^{2}
\end{equation}
We have used (+, -, -, -) signature convention and a
Friedmann-Robertson-Walker line element given by
\begin{equation}
ds^{2}=dt^{2}-\frac{d\sigma^{2}e^{g}}{\left( 1+Br^{2}\right)^{2}}
\end{equation}
where $d \sigma^{2}$ is the three-dimensional Euclidian line element
and $A=8\pi G/c^{2}$, $B=k/4a^{2}$ and with $ k=0$ , $k=1$ and
$k=-1$. We also have $a^{2}$ as a constant.
\newline
In this paper we pay attention to our spatial flat solution,  $B=0$.
In this case the field is given by
\begin{equation}
\phi =\frac{\in mt}{\sqrt{3A}}+b
\end{equation}
with $\in =\pm 1$ and $b$ is an arbitrary constant.
\newpage
The cosmological constant obeys the condition
\begin{equation}
\Lambda =-\frac{m^{2}}{3}=-\frac{1}{3}( \frac{cM}{\hbar})^2,
\end{equation}
a negative value, associated with the Planck's constant, the speed
of light and a scalar particle of mass $M$.
\newline
The corresponding line element is
\begin{equation}
ds^{2}=dt^{2}-d\sigma ^{2}e^{[-2\in
mb(\sqrt{\frac{A}{3}})t-\frac{m^{2}}{3} t^{2}]}.
\end{equation}
The universe is in an expansive phase. For $t>0$ we choose $\in =-1$
and $b>0$. Then, we rewrite (2.8) as
\begin{equation}
ds^{2}=dt^{2}-d\sigma
^{2}e^{[2mb(\sqrt{\frac{A}{3}})t-\frac{m^{2}}{3} t^{2}]}.
\end{equation}
For $t=-\tau<0$ we choose $\in =+1$, and $b>0$. Then, we rewrite
(2.8) as
\begin{equation}
ds^{2}=d\tau^{2}-d\sigma
^{2}e^{[2mb(\sqrt{\frac{A}{3}})\tau-\frac{m^{2}}{3} \tau^{2}]}.
\end{equation}

In this paper we use another convention to the Riemann tensor, as
follows,
\begin{equation}
R^{\alpha}{}_{\mu \sigma \nu }=-\partial_{\nu} \Gamma_{\mu \sigma
}^{\alpha}+\partial_{\sigma}\Gamma_{\mu \nu}^{\alpha}-\Gamma_{\mu
\sigma}^{\eta}\Gamma_{n \nu}^{\alpha}+\Gamma_{\mu \nu}^{\eta}\Gamma
_{\sigma \eta}^{\alpha}
\end{equation}
which implies
\begin{equation}
R_{\mu \nu }-\frac{1}{2}g_{\mu \nu }R-\Lambda g_{\mu \nu}=\frac{8\pi
G}{ c^{2}}T_{\mu \nu}.
\end{equation}
The motion will be simpler in a Fermi-Walker transported tetrad
basis.
\newline
Let us consider the connection between the tetrad and the local
metric tensor
\begin{equation}
g_{\lambda\pi}=E_{\lambda}^{(\mathbf{A})}E_{\pi}^{(\mathbf{B})}\eta_{(\mathbf{A})(\mathbf{B})},
\end{equation}
where $ \eta_{(\mathbf{A})(\mathbf{B})}$ and $
E_{\lambda}^{(\mathbf{A})}$ are the Lorentzian metric and tetrad
components, respectively.
\newpage
From (2.8) we have
\begin{equation}
E_{0}^{(\mathbf{0})}=1,
\end{equation}
\begin{equation}
E_{1}^{(\mathbf{1})}=E_{2}^{(\mathbf{2})}=E_{3}^{(\mathbf{3})}=e^{[-\in
mb(\sqrt{\frac{A}{3}})t-\frac{m^{2}}{6} t^{2}]}.
\end{equation}
We now write the $1$-form
\begin{equation}
\theta^{(\mathbf{A})}= dx^{\lambda} E_{\lambda}^{(\mathbf{A})}.
\end{equation}
By exterior derivatives of (2.16) and using the Cartan's second
structure equation, we obtain
\begin{eqnarray}
 \nonumber R^{(\mathbf{1})}{_{(\mathbf{0})(\mathbf{0})(\mathbf{1})}}=R^{(\mathbf{2})}{_{(\mathbf{0})(\mathbf{0})(\mathbf{2})}}=\\
 \nonumber
 =R^{(\mathbf{3})}{_{(\mathbf{0})(\mathbf{0})(\mathbf{3})}}=-\frac{m^{2}}{3}+\frac{1}{2}[-2\in
mb(\sqrt{\frac{A}{3}})-\frac{m^{2}}{3}t]^{2}.\\
\end{eqnarray}
Let us present the Jacobi equation in a Fermi-Walker transported
tetrad basis \cite{5},
\begin{equation}
\frac{d^2Z^{(\mathbf{A})}}{d\tau
^2}+R^{(\mathbf{A})}{_{(\mathbf{0})(\mathbf{C})(\mathbf{0})}}Z^{(\mathbf{C})}=0.
\end{equation}
Substituting (2.17) in (2.18) we obtain
\begin{equation}
\frac{d^2Z^{(\mathbf{A})}}{d\tau
^2}=\{-\frac{m^{2}}{3}+\frac{1}{2}[-2\in
mb(\sqrt{\frac{A}{3}})-\frac{m^{2}}{3}t]^{2}\}Z^{(\mathbf{A})},
\end{equation}
with $A=(1,2,3)$.
\newline
We can rewrite (2.19) as follows
\begin{equation}
\frac{d^2Z^{(\mathbf{A})}}{d\tau
^2}=\{-\frac{m^{2}}{3}-2bm^{3}\frac{1}{3}(\sqrt{\frac{A}{3}})t
+\frac{2m^{2}}{3}b^{2}A+\frac{m^{4}}{18}t^{2}\}Z^{(\mathbf{A})}.
\end{equation}
The Jacobi equation will be appropriate to show the relative
acceleration between two particles  if we do not have to consider
the metric deformation by particles. For the primordial universe
(2.8), we have from (2.19) or (2.20) that two  massive scalar
particles in two geodesics close to each other feel two primordial
forces, one attractive (dark matter) and another repulsive (dark
energy) both increasing with distance. The same scalar particle will
be responsible for the two metric forces which, conveniently, we
have identified as dark matter and dark energy. From the
gravitational point of view, the creation of other types of matter
by the universe generates three competitive forces, the two
primordial forces above presented, and another which, for galaxies,
can be expressed by the Newtonian gravity. Inside and outside the
galaxies, the resulting force is the sum of these three forces. A
correct dynamic description of one or more stars in a galaxy,
depends on a set of information about galaxy evolution. Elliptical
and spiral galaxies, as well as clusters of galaxies have different
dynamics and different evolution processes. The Newtonian gravity is
very important to describe the galaxies dynamics but it is not
enough. Physicists and astronomers have concluded that the Newtonian
gravity only is not sufficient to describe the galaxies dynamics.
They believe in the existence of a second attractive force (dark
matter) which, in association with the Newtonian gravity, governs
the star dynamics. They also believe in the existence of a repulsive
force (dark energy) responsible for the expansion on large scale. We
believe that the presence of the two primordial forces together with
the Newtonian force, can describe the galaxies's behavior. Inside
and outside the galaxies the resulting force is the sum of the three
forces. The value of the constant $b$ in (2.6) and (2.8) could be
fixed by experimental records of galaxies (dark matter) or
cosmological expansion (dark energy) or both. Modifications in the
stars motion in galaxies, can be made by appropriate adjustments in
the constant $b$. It is possible  that $b$ is a new constant of
nature, as well as the mass $M$ of the scalar particle.
\newline
The mass of the scalar particle can be estimated by astronomic
measurements of cosmological constant. From (2.7) and (2.19) we
conclude that $\Lambda$ is associated with an attractive force, for
us conveniently identified as dark matter, and a repulsive force,
identified as dark energy. The second term in the second member of
(2.19) is positive and, therefore, identified as dark energy. Notice
that the dark energy term is a function of time, of the constant
$b$, and also of the term identified as the dark matter. We noticed
in (2.20) that the constants $m$ and $b$ are present in the terms
associated with the dark matter (attractive force) and to the terms
associated with the dark energy (repulsive force). Then, $\Lambda$
is present in the dark matter and in the dark energy. In other
words, it is not possible to separate them, because dark matter and
dark energy are scalar particle effects. Originally, the
cosmological constant was associated with a repulsive force so that
the estimate given in the following is associated with a large scale
expansion. It can be slightly different from a realistic and
definitive value, but our objective is to point that we cannot
detect the scalar particle, or, at best, we have a very low
probability of doing so.
\newline
Using the experimental limit for the constant  $\Lambda$ in (2.7)
\cite{6}, it is possible to obtain  a superior limit for the mass
 of the scalar particle. The cosmological constant was estimated as
\begin{equation}
\Lambda<{10}^{-54}cm^{-2}.
\end{equation}
Using it and (2.7) we obtain
\begin{equation}
M<(6).{10}^{-65}g.
\end{equation}
where (2.35) was used. There is another limit for the cosmological
constant \cite{3} given by
\begin{equation}
\Lambda{Lp}^2<{10}^{-123},
\end{equation}
or
\begin{equation}
\Lambda<{10}^{-57}cm^{-2}.
\end{equation}
Using it and (2.7) we obtain
\begin{equation}
M<(1.9).{10}^{-66}g.
\end{equation}
The relationship between the electron rest mass and the mass of the
scalar particle is approximately given by
\begin{equation}
m_{e}\sim (4.79).{10}^{38}M.
\end{equation}
The universe expansion can be calculated. In other words, it is
possible to calculate the starting point of the universe
contraction. Using (2.33) in (2.8) we obtain
\begin{equation}
t\sim\frac{1}{\sqrt{-\Lambda}}.
\end{equation}
Notice that we have assumed $c=1$ in (2.8).  Therefore, for numeric
results involving time, we regain $ct$, so that
\begin{equation}
t\sim{10}^{27}\frac{cm}{c}\sim(3.33){10}^{9}years,
\end{equation}
From (2.24) and (2.27) we have
\begin{equation}
t\sim(3.162){10}^{28}\frac{cm}{c}\sim(33.4){10}^{9}years.
\end{equation}
Note that (2.28) is incompatible with the geological data of the
Earth. If (2.29) is  a good estimate, it will be almost impossible
to detect the scalar particle. Its influence will be predominantly
gravitational and it is given by (2.8). Consequently, for many
classical situations as, for instance, the solar system dynamics,
the effect on the ordinary matter would be insignificant. For this
condition we consider only the ordinary matter in the Einstein's
equation. The scalar particle can be very important for galaxies,
clusters of galaxies, and large structures. It is necessary an
investigation to evaluate the influence of an intense gravitational
field generated by a classical black hole geometry in the primordial
scalar particle.
\newline
The primordial universe (2.8) starts with scalar particles and is
non-singular  at $t=0$. It is an expansible universe if the
curvature ${R^{(A)}}_{(A)}$ obeys a simple inequality. With the time
evolution, other types of matter were created and complex
interactions among particles are checked every day. Analytical
solutions of (2.10) with the inclusion of other fields are very
difficult. However, as the influence of the primordial universe
(2.8) could have been very important in the past and can be very
important in the present, it is reasonable to suppose that other
primordial particles are ghosts, so that (2.8) is a consequence of
the scalar particles only. In other words, the momentum-energy
tensors of other primordial fields have not contributed to the
curvature of the primordial universe in the past nor in the present,
although such particles interact with all that, playing an important
part in the evolution of the universe as well as in the creation of
the ordinary matter. We recall that in the cosmological models,
metrics as the Friedmann-Robertson-Walker are important to the
initial large structure formation as well as to the universe
evolution. But, gradually each local distribution of matter will be
more and more important and the effect of all distributions of
matter in the universe is represented by a moment-energy tensor of a
fluid in the Einstein's equation for a Friedmann-Robertson-Walker
metric. However, if (2.8) is responsible for the dark matter and the
dark energy, we will have a different situation. In this case, (2.8)
would determine the evolution of the universe in the past and in the
present, and due to the mass estimate of the scalar particle, its
interaction with other particles would be predominantly
gravitational.
\newline
It is important to notice the presence of two different time scales,
one associated with a local distribution, as well as with a large
scale structure of ordinary matter, and another associated with the
cosmological time of (2.8). The embedding of (2.8) and of a
classical metric in an $n$ dimensional flat space is a possible
strategy to consider the gravitational interaction between the
scalar particles and the ordinary matter.
\newline
The primordial universe (2.8) is non-singular at $t=0$. It is
cyclical and eternal, and could have different cycles. Although it
is not the only possibility, a negative curvature is the simplest
mechanism for an expansive universe and it will be considered.
\newline
For (2.8) the curvature is given by
\begin{eqnarray}
 \nonumber{R^{(A)}}_{(A)} =\\
 \nonumber
 =-2(\frac{cM}{\hbar})^2 +6[-2\in
\frac{cM}{\hbar}b(\sqrt{\frac{A}{3}})-\frac{(\frac{cM}{\hbar})^2 }{3}t]^{2}.\\
\end{eqnarray}
We have at $t=0$
\begin{eqnarray}
 \nonumber{R^{(A)}}_{(A)}(t=0) =\\
 \nonumber
 =2(\frac{cM}{\hbar})^2[-1+4b^2.A],\\
\end{eqnarray}
which is a finite curvature. We consider an initial negative
curvature ${R^{(A)}}_{(A)}$ as the simplest condition for the
primordial expansive universe
\begin{eqnarray}
 \nonumber{R^{(A)}}_{(A)}(t=0) <0,\\
\end{eqnarray}
so that
\begin{eqnarray}
\| b\|<\frac{1}{2\sqrt{A}},
\end{eqnarray}
or
\begin{eqnarray}
\| b\|<\frac{c}{2\sqrt{8{\pi}G}},
\end{eqnarray}
or
\begin{eqnarray}
\| b\|<{10}^{15}\sqrt{\frac{g}{cm}},
\end{eqnarray}
where (2.35) is a superior limit for $b.$ We have obtained two
superior limits for the mass of the scalar particle and for the
constant $b$ given by (2.22) and (2.35), respectively. Note that $b$
and $M$ can be two new constants of nature. For (2.24), $M$ will be
given by (2.25), smaller than (2.22), reinforcing the previous
conclusion that it is very difficult to detect this scalar particle.
\newline
Note that our choice of an initial negative curvature, as the
expansion mechanism,  imposed a superior limit for the constant b.
However, other mechanisms are possible, so that the constant b can
assume another limit, compatible with experimental results.
\newline
In section $3$ we use the first quantization to interactions among
primordial scalar particles, primordial ghost fermions, and ordinary
fermions.
\renewcommand{\theequation}{\thesection.\arabic{equation}}
\section{\bf The Dirac's Equation}
\setcounter{equation}{0}
 $         $
In this section we consider the Dirac's equation and show, using the
PCT theorem, the particle-antiparticle asymmetry  in a primordial
universe as (2.8).
\newline
To preserve the solution (2.8) and consequently the Einstein's
equation, it is reasonable to suppose that other primordial fields
are ghosts, so that (2.8) is a consequence of the scalar particles
only. In other words, the momentum-energy tensors of other
primordial fields have not contributed and still do not contribute
today to the curvature of the primordial universe, although such
particles interact with all that, playing an important part in the
evolution of the universe as well as in the creation of the ordinary
matter. These other primordial fields can be identified as some
quantum vacuum constituents.
\newline
In the following, we consider the Dirac momentum-energy tensor
\cite{7}, \cite{8},\cite{9},
\begin{eqnarray}
 \nonumber T_{(A)(B)}=i\{\bar{\psi}\gamma_{(A)}\nabla_{(B)}\psi+\bar{\psi}\gamma_{(B)}\nabla_{(A)}\psi
 \nonumber-\nabla_{(A)}\bar{\psi}\gamma_{(B)}\psi-\nabla_{(B)}\bar{\psi}\gamma_{(A)}\psi\}.\\
\end{eqnarray}
For ghost fields, the momentum-energy tensor obeys the following
condition \cite{10},
\begin{eqnarray}
 T_{(A)(B)}=0.
\end{eqnarray}
The Dirac's equation in a curved space is given by
\begin{equation}
 i\gamma^{(\mathbf{D})}(\partial_{(D)}
 -\Gamma_{(D)})\psi-\bar{m}\psi=0,
\end{equation}
where
\begin{equation}
 \Gamma_{(D)}=-ieA_{(D)}I-\frac{1}{4}\gamma_{(A) (B)
 (C)}\gamma^{(\mathbf{A})}\gamma^{(\mathbf{B})},
\end{equation}
and the $\gamma_{(A) (B) (C)}$  are the Ricci's rotation
coefficients.
\newline
For two spaces, one flat and the other curved, the Dirac's matrices
are related by
\begin{equation}
\gamma^{(\mathbf{A})}=E_{\lambda}^{(\mathbf{A})}(x)\gamma^{\lambda}(x).
\end{equation}
In this paper we use the following representation
\begin{equation}
 \gamma^{(0)}=\left(%
\begin{array}{cc}
  I & 0 \\
  0 & -I \\
\end{array}%
\right)
\end{equation}
\begin{equation}
 \gamma^{(k)}=\left(%
\begin{array}{cc}
  0 & \sigma^{k} \\
  -\sigma^{k} & 0 \\
\end{array}%
\right)
\end{equation}
\begin{equation}
\gamma^{\mathbf{k}}=\frac{1}{R}\gamma^{(\mathbf{k})},
\end{equation}
where from (2.15)
\begin{equation}
R=e^{-2[\in mb(\sqrt{\frac{A}{3}})t+\frac{m^{2}}{6} t^{2}]}.
\end{equation}
For (2.8) we have
\begin{equation}
 \Gamma_{(0)}=-ieA_{(0)}I,
\end{equation}
\begin{equation}
 \Gamma_{(k)}=-ieA_{(k)}I+\frac{\dot{R}}{2R}\gamma^{(\mathbf{k})}\gamma^{(\mathbf{0})}.
\end{equation}
Substituting (3.3) in (3.2) we obtain
\begin{equation}
 \gamma^{(\mathbf{D})}[i\partial_{(D)}-eA_{(D)}+i\frac{1}{4}\gamma_{(A) (B) (D)}\gamma^{(\mathbf{A})}\gamma^{(\mathbf{B})})]\psi-\bar{m}\psi=0,
\end{equation}
where $\bar{m}$ is a fermion mass.
\newline
By a charge conjugation we obtain
\begin{equation}
 \gamma^{(\mathbf{D})}[i\partial_{(D)}+eA_{(D)}+i\frac{1}{4}\gamma_{(A) (B)
 (D)}\gamma^{(\mathbf{A})}\gamma^{(\mathbf{B})})]\psi_{(c)}-\bar{m}\psi_{(c)}=0.
\end{equation}
Using (3.12) into (3.1), we verify the condition (3.2) if
\begin{equation}
A_{(k)}=0,
\end{equation}
\begin{equation}
A_{(0)}=0,
\end{equation}
and
\begin{eqnarray}
 \psi=\psi(t),
\end{eqnarray}
where $\psi(t)$ is a function of time only.
\newline
Explicitly
\begin{equation}
\partial_{t}\psi+\frac{3\dot{R}}{2R}\psi+i\bar{m}\gamma^{(\mathbf{0})}\psi=0,
\end{equation}
\begin{equation}
\partial_{t}\psi_{(c)}+\frac{3\dot{R}}{2R}\psi_{(c)}+i\bar{m}\gamma^{(\mathbf{0})}\psi_{(c)}=0,
\end{equation}
In the following we present the difference between $\psi_{l}$ and
$\psi^{(r)}$.  $\psi^{(r)}$ is one of the four linearly independent
solutions of the Dirac's equation in the form of column matrices,
and $\psi_{l}$ is an element of one of these four matrices. We use
similar considerations for the charge conjugate solutions.
\newline
Substituting (3.16) and (3.9) in (3.12)
\begin{equation}
\partial_{t}\psi_{j}+\{i\bar{m}c^2 +\frac{3\dot{R}}{2R}\}\psi_{j}=0,
\end{equation}
\begin{equation}
\partial_{t}\psi_{l}+\{-i\bar{m}c^2
+\frac{3\dot{R}}{2R}\}\psi_{l}=0.
\end{equation}
Explicitly
\begin{equation}
\partial_{t}\psi_{j}+\{i\bar{m}c^2-3\in mb\sqrt{\frac{A}{3}}-m^2t\}\psi_{j}=0,
\end{equation}
\begin{equation}
\partial_{t}\psi_{l}+\{-i\bar{m}c^2-3\in mb\sqrt{\frac{A}{3}}-m^2t\}\psi_{l}=0,
\end{equation}
where $j=1,2$ and $l=3,4$.
\newline
By integration of (3.19) and (3.20) we have
\begin{equation}
\psi_{j}=e^{-i\bar{m}c^2t}\exp{(-\frac{3\dot{R}}{2R})}c_{j},
\end{equation}
\begin{equation}
\psi_{k}=e^{i\bar{m}c^2t}\exp{(-\frac{3\dot{R}}{2R})}c_{k},
\end{equation}
where $c_{j}$ and $ c_{k}$ are constants.
\newline
The four linearly independent solutions  are given by
\begin{equation}
\psi^{(r)}=e^{-i\bar{m}c^2t}\exp{(-\frac{3\dot{R}}{2R})}u^{(r)}
\end{equation}
\begin{equation}
\psi^{(s)}=e^{i\bar{m}c^2t}\exp{(-\frac{3\dot{R}}{2R})}u^{(s)}
\end{equation}
Explicitly we have
\begin{equation}
\psi^{(r)}=e^{-i\bar{m}c^2t}e^{[3\in
mb\sqrt{\frac{A}{3}}t+\frac{m^2}{2}t^2]}u^{(r)},
\end{equation}
\begin{equation}
\psi^{(s)}=e^{i\bar{m}c^2t}e^{[3\in
mb\sqrt{\frac{A}{3}}t+\frac{m^2}{2}t^2]}u^{(s)},
\end{equation}
where
\begin{equation}
     u^{(r)}=\left(%
              \begin{array}{c}
                \delta_{1r} \\
                \delta_{2r} \\
                0 \\
                0 \\
              \end{array}%
            \right)
\end{equation}
\begin{equation}
u^{(s)}=\left(%
        \begin{array}{c}
          0 \\
          0 \\
          \delta_{3s} \\
          \delta_{4s} \\
        \end{array}%
      \right)
\end{equation}
where $r=1,2$ and $s=3,4$.
\newline
Note that the wave function (3.16), for the metric (2.8), is given
by
\begin{eqnarray}
\nonumber\psi=\psi(t)=e^{[3\in
mb\sqrt{\frac{A}{3}}t+\frac{m^2}{2}t^2]}[\sum_{r}c_{r}e^{-i\bar{m}c^2t}u^{(r)}\\
\nonumber+\sum_{s}d_{s}e^{i\bar{m}c^2t}u^{(s)}],\\
\end{eqnarray}
with
\begin{equation}
\|c_{1}\|^{2}+\|c_{2}\|^{2}=\|d_{3}\|^{2}+\|d_{4}\|^{2},
\end{equation}
obtained from the condition
\begin{equation}
\bar{\psi}\psi=0,
\end{equation}
for ghost fermions in (2.8).
\newline
Note that (3.33) is also true if (3.16) is given by one of the four
solutions (3.27) and (3.28), or a linear combination of two
solutions, one from (3.27) and the other from (3.28). The above is
also true for ghost neutrinos.
\newline
For ghost neutrinos we have
\begin{equation}
\partial_{t}\psi_{j}-\{3\in mb\sqrt{\frac{A}{3}}+m^2t\}\psi_{j}=0,
\end{equation}
\begin{equation}
\partial_{t}\psi_{l}-\{3\in mb\sqrt{\frac{A}{3}}+m^2t\}\psi_{l}=0,
\end{equation}
where $j=1,2$ and $l=3,4$.
\newline
The solutions are given by
\begin{equation}
\psi^{(r)}=e^{[3\in mb\sqrt{\frac{A}{3}}t+\frac{m^2}{2}t^2]}u^{(r)}
\end{equation}
\begin{equation}
\psi^{(s)}=e^{[3\in mb\sqrt{\frac{A}{3}}t+\frac{m^2}{2}t^2]}u^{(s)}
\end{equation}
where $r=1,2$ and $s=3,4$.
\newline
Now we will define the following wave functions
\begin{equation}
\Phi^{(s)}=\exp{(\frac{3\dot{R}}{2R})}\psi^{(s)},
\end{equation}
\begin{equation}
\Phi^{(r)}=\exp{(\frac{3\dot{R}}{2R})}\psi^{(r)},
\end{equation}
so that
\begin{equation}
\Phi^{(r)}=e^{-i\bar{m}c^2t}u^{(r)},
\end{equation}
\begin{equation}
\Phi^{(s)}=e^{i\bar{m}c^2t}u^{(s)}.
\end{equation}
The wave functions (3.40) and (3.41) are the four linearly
independent solutions of the Dirac's equation for a rest particle in
the Minkowski space-time
\begin{equation}
\partial_{t}\Phi+i\bar{m}c^{2}\gamma^{(\mathbf{0})}\Phi=0.
\end{equation}
Substituting (3.38) and (3.39) in (3.42) we obtain (3.17). For a
Lorentz's boost applied to (3.40) and (3.41) we obtain plane wave
functions for a free particle. This does not imply, as a consequence
of (3.38) and (3.39), that the boost is a correct procedure for
$\psi^{(r)}$ and $ \psi^{(s)}$. In other words, the new functions,
obtained from $\psi^{(r)}$ and $ \psi^{(s)}$   by a boost, are not
solutions of the Dirac's equation.
\newline
Although the calculus was based on (2.8), the expansible initial
phase of (2.8), with $t>0$ and $\epsilon=-1$, is given by (2.9). It
is interesting to rewrite the wave functions as follows
\begin{equation}
\psi^{(r)}=e^{-i\bar{m}c^2t}e^{[
-3mb\sqrt{\frac{A}{3}}t+\frac{m^2}{2}t^2]}u^{(r)}
\end{equation}
\begin{equation}
\psi^{(s)}=e^{i\bar{m}c^2t}e^{[
-3mb\sqrt{\frac{A}{3}}t+\frac{m^2}{2}t^2]}u^{(s)}
\end{equation}
for a massive fermion, and
\begin{equation}
\psi^{(r)}=e^{[-3mb\sqrt{\frac{A}{3}}t+\frac{m^2}{2}t^2]}u^{(r)}
\end{equation}
\begin{equation}
\psi^{(s)}=e^{[-3mb\sqrt{\frac{A}{3}}t+\frac{m^2}{2}t^2]}u^{(s)}
\end{equation}
for a neutrino.
\newline
The simplified form of the wave functions is a consequence of the
symmetries in the primordial universe, as well as the ghost
particles hypothesis for other primordial particles.
\newline
Notice that the cosmological time $t$ should be used in the Dirac's
equation for the primordial phase of the universe. In the present
phase we should use it in the  Dirac's equation when we consider the
evolution of ghost fermions only. For interactions involving
ordinary matter, the time in the Dirac's equation will be another
time associated with the ordinary matter.
\newline
Strong reflection ($x'^{(\mathbf{D})} \rightarrow -x^{(\mathbf{D})}$
) followed by Hermitian conjugation is equivalent to time
reflection, charge conjugation, and parity ($PCT$), if the product
of the three phases is chosen to be one \cite{11}.
\newline
The $PCT$ theorem can be written as follows,
\begin{equation}
\psi'(x'^{(\mathbf{D})})=i\gamma^{(5)}\psi(x^{(\mathbf{D})}),
\end{equation}
with
\begin{equation}
x'^{(\mathbf{D})}=-x^{(\mathbf{D})}.
\end{equation}
Let us consider the Dirac's equation as follows
\begin{equation}
 \gamma^{(\mathbf{D})}\partial_{(D)}\psi+i\bar{m}c^2\psi+
 \frac{1}{4}\gamma_{(A) (B)
 (D)}\gamma^{(\mathbf{D})}\gamma^{(\mathbf{A})}\gamma^{(\mathbf{B})}\psi=0.
\end{equation}
Multiplying it by $\gamma^{(\mathbf{5})}$ from the left and using
the $\gamma's$  properties
\begin{equation}
 -\gamma^{(\mathbf{D})}\partial_{(D)}\gamma^{(\mathbf{5})}\psi(x^{(\mathbf{D})})+i\bar{m}c^2\gamma^{(\mathbf{5})}\psi-
 \frac{1}{4}\gamma_{(A) (B)
 (D)}\gamma^{(\mathbf{D})}\gamma^{(\mathbf{A})}\gamma^{(\mathbf{B})}\gamma^{(\mathbf{5})}\psi=0.
\end{equation}
But
\begin{equation}
-\partial_{(D)}\gamma^{(\mathbf{5})}\psi(x^{(\mathbf{D})})=\partial_{(D')}\gamma^{(\mathbf{5})}\psi(x^{(\mathbf{D})}).
\end{equation}
Substituting in (3.50) we obtain
\begin{equation}
\nonumber
\gamma^{(\mathbf{D})}\partial_{(D')}\gamma^{(\mathbf{5})}\psi(x^{(\mathbf{C})})\\
\nonumber+i\bar{m}c^2\gamma^{(\mathbf{5})}\psi-\frac{1}{4}\gamma_{(A) (B)\\
\nonumber(D)}(x^{(\mathbf{C})})\gamma^{(\mathbf{D})}\gamma^{(\mathbf{A})}\gamma^{(\mathbf{B})}\gamma^{(\mathbf{5})}\psi=0.\\
\end{equation}
 For the coordinate $x'^{(\mathbf{D})}$ the covariance of the Dirac's equation implies
\begin{equation}
 \gamma^{(\mathbf{D})}\partial_{(D')}\psi'(x'^{(\mathbf{C})} )+i\bar{m}c^2\psi'(x'^{(\mathbf{C})})+
 \frac{1}{4}\gamma_{(A) (B)
 (D)}(x'^{(\mathbf{C})})\gamma^{(\mathbf{D})}\gamma^{(\mathbf{A})}\gamma^{(\mathbf{B})}\psi'(x'^{(\mathbf{C})} )=0.
\end{equation}
For (2.8) the equation (3.53) will be
\begin{equation}
\partial_{t'}\psi(t')+\frac{3R'}{2R}\psi(t')+i\bar{m}\gamma^{(\mathbf{0})}\psi(t')=0,
\end{equation}
or
\begin{equation}
\partial_{t'}\psi'(t')+\{-3\in
mb\sqrt{\frac{A}{3}}-m^2t'\}\psi'(t')+i\bar{m}\gamma^{(\mathbf{0})}\psi(t')=0,
\end{equation}
with solutions
\begin{equation}
\psi'^{(r)}(t')=e^{-i\bar{m}c^2t'}e^{[3\in
mb\sqrt{\frac{A}{3}}t'+\frac{m^2}{2}t'^2]}u^{(r)}
\end{equation}
\begin{equation}
\psi'^{(s)}(t')=e^{i\bar{m}c^2t'}e^{[3\in
mb\sqrt{\frac{A}{3}}t'+\frac{m^2}{2}t'^2]}u^{(s)}
\end{equation}
Substituting (3.48) in (3.56) and in (3.57) we obtain
\begin{equation}
\psi'^{(r)}(t')=\psi'^{(r)}(-t) =e^{i\bar{m}c^2t}e^{[-3\in
mb\sqrt{\frac{A}{3}}t+\frac{m^2}{2}t^2]}u^{(r)},
\end{equation}
\begin{equation}
\psi'^{(s)}(t')=\psi'^{(s)}(-t)=e^{-i\bar{m}c^2t}e^{[-3\in
mb\sqrt{\frac{A}{3}}t+\frac{m^2}{2}t^2]}u^{(s)}.
\end{equation}
For (2.8) the equation (3.50) (or (3.52) ) will be
\begin{equation}
 -\partial_{t}\gamma^{(\mathbf{5})}\psi(t)+\{3\in
mb\sqrt{\frac{A}{3}}+m^2t\}\gamma^{(\mathbf{5})}\psi(t)
+i\bar{m}c^2\gamma^{(\mathbf{0})}\gamma^{(\mathbf{5})}\psi(t)=0,
\end{equation}
with solutions
\begin{equation}
\gamma^{(\mathbf{5})}\psi^{(r)}(t)=e^{i\bar{m}c^2t}e^{[3\in
mb\sqrt{\frac{A}{3}}t+\frac{m^2}{2}t^2]}u^{(r)},
\end{equation}
\begin{equation}
\gamma^{(\mathbf{5})}\psi^{(s)}(t)=e^{-i\bar{m}c^2t}e^{[3\in
mb\sqrt{\frac{A}{3}}t+\frac{m^2}{2}t^2]}u^{(s)}.
\end{equation}
We conclude that (3.58), (3.59), (3.61) and (3.62) do not obey the
$PCT$ theorem (3.47) and (3.48). The $PCT$ theorem is violated in
the primordial phase of the universe.
\newline
Note that $t'$ is a negative time. As in the obtention of (2.10),
make $t'=-\tau$, where $\tau$ is a positive time. It is convenient
to avoid a confusing notation. Then, for $\in =+1$, and $b>0$, as in
(2.10), the solutions (3.56) and (3.57) assume the following form
\begin{equation}
\psi'^{(r)}(t')=e^{i\bar{m}c^2\tau}e^{[
-3mb\sqrt{\frac{A}{3}}\tau+\frac{m^2}{2}{\tau}^2]}u^{(r)},
\end{equation}
\begin{equation}
\psi'^{(s)}(t')=e^{-i\bar{m}c^2\tau}e^{[
-3mb\sqrt{\frac{A}{3}}\tau+\frac{m^2}{2}{\tau}^2]}u^{(s)}.
\end{equation}
We conclude that (3.63) and (3.64) are calculated in the universe
(2.10).
\newline
The solutions (3.61) and (3.62) live in our universe (2.9), where
$t>0$, $\in =-1$, and $b>0$. In (2.9) they assume the following
forms
\begin{equation}
\gamma^{(\mathbf{5})}\psi^{(r)}(t)=e^{i\bar{m}c^2t}e^{[
-3mb\sqrt{\frac{A}{3}}t+\frac{m^2}{2}t^2]}u^{(r)},
\end{equation}
\begin{equation}
\gamma^{(\mathbf{5})}\psi^{(s)}(t)=e^{-i\bar{m}c^2t}e^{[
-3mb\sqrt{\frac{A}{3}}t+\frac{m^2}{2}t^2]}u^{(s)}.
\end{equation}
The solutions (3.63), (3.64), (3.65) and (3.66) satisfy the Dirac's
equation and obey the $PCT$ theorem.
\section{Concluding Remarks}
  $              $
The embedding in an n-dimensional flat space of one metric only  is
well known. The embedding of (2.8) and of a classical metric in an
n-dimensional flat space is a possible strategy to consider the
gravitational interaction between the scalar particle and the
ordinary matter.
\newline
The Jacobi  equation (2.20) eliminates fictitious forces and
identifies attractive (dark matter) and repulsive (dark energy)
forces as a scalar particle effect.
\newline
Strong reflection ($x^{(\mathbf{D})} \rightarrow -x^{(\mathbf{D})}$)
followed by Hermitian conjugation is equivalent to time reflection,
charge conjugation, and parity ($PCT$), \cite{11}.
\newline
The solutions (3.40) and (3.41) of the Dirac's equation (3.42) for
massive fermions obey the $PCT$ theorem. The same is not true for
the solutions (3.25) and (3.26) of the Dirac's equation (3.17).
Then, the $PCT$ theorem is violated in the primordial phase of the
universe. In this phase the particle-antiparticle symmetry is
incompatible with (2.8). The solutions of the Dirac's equation
present a real and initial expansible factor. If we use the $PCT$
theorem  in the anti-fermion (or fermion) wave functions, we will
have an asymmetry between the initial expansible factors due to
$t\rightarrow -t$. Consequently, for creation and annihilation
processes, fermions and anti-fermions have different probabilities.
This could give the necessary and quantitative initial difference
between particles and antiparticles to explain the observed
asymmetry. If we consider that (2.8) represents two expansible
parallel universes, one of particles, with $t>0$, $\epsilon=-1$,
$b>0$, described by (2.9), and the other of anti-particles, with
$t'=-t=-\tau<0$, $\epsilon=+1$, and $b>0$, described by (2.10), then
the $PCT$ theorem will be obeyed if $\gamma^{(\mathbf{5})}\psi(t)$
and $\psi'(t')$ are calculated in (2.9) and (2.10), respectively.
For the particle parallel universe (2.9), the presence of
anti-particles will be possible by processes involving
energy-momentum conservation, as a pair creation. We interpret the
presence of anti-particles in the parallel universe (2.9) as a
capture from the parallel universe (2.10). Processes involving
energy-momentum conservation, as a pair creation, can be a bridge
between the two primordial universes. We believe that future
attempts at anti-hydrogen trapping will be successful \cite{12}.
From our above hypotheses, anti-hydrogen trapping can be thought as
a capture from the antiparticle parallel universe (2.10). This
allows us to consider the possibility of a temporary hydrogen
trapping in the parallel universe (2.10) as a symmetric process. It
is a conjecture.
\newline
If (2.8)  correctly describes the dark matter and the dark energy as
an effect of the primordial scalar particle, the Einstein's equation
needs to be preserved. This implies in considering other primordial
particles as being ghosts. If we preserve the Dirac's equation and
the $PCT$ theorem in the primordial phase,  we will need to consider
the existence of the two parallel and expansible primordial
universes (2.9) and (2.10).
\newline
Notice that the cosmological time $t$ should be used in the Dirac's
equation for the primordial phase of the universe. In the present
phase we should use it in the  Dirac's equation when we consider the
evolution of ghost fermions only. For interactions involving
ordinary matter, the time in the Dirac's equation will be another
time, associated with the ordinary matter. In this case the metric
(2.8) can be considered as a small background field, and the PCT
theorem can be obeyed, but embedding will be necessary for a more
realistic approach.

\end{document}